\documentclass[12pt]{amsart}
\usepackage{graphicx}
\usepackage{booktabs}
\usepackage{amsmath,amssymb,amsthm,graphicx,url,etoolbox}
\usepackage{natbib,fullpage}

\usepackage{times}
\usepackage{pdflscape}

\newcommand{\pandc}{packing and cracking}
\newcommand{\spc}{SPC}

\newcommand{\bl}{\boldsymbol{\ell}}

\newcommand{\bnl}{\boldsymbol{\ell}^*}

\theoremstyle{definition}
\newtheorem{theorem}{Theorem}

\newtheorem{example}[theorem]{Example}

\newcommand{\lambdaeg}[1]{{{EG}}}
\newcommand{\eg}{{\lambdaeg{1}}}

\newcommand{\mm}{MM}
\newcommand{\bias}{{GK-Bias}}

\newcommand{\dec}{{Dec}}

\newcommand*\rot{\rotatebox{90}}

\makeatletter
\g@addto@macro\bfseries{\boldmath}
\makeatother

\title{Simulated packing and cracking}

\ifdef{\SUBMIT}{}{
\author{Jeffrey S. Buzas}
\address{Department of Mathematics \& Statistics, University of Vermont,16 Colchester Ave., Burlington, VT 05401, USA}
\email{jeff.buzas@uvm.edu}

\author{Gregory S. Warrington}
\address{Department of Mathematics \& Statistics, University of Vermont,16 Colchester Ave., Burlington, VT 05401, USA}
\email{gswarrin@uvm.edu}

\date{}
}

\begin{document} 



\begin{abstract}
We introduce simulated packing and cracking as a technique for
evaluating partisan-gerrymandering measures. We apply it to historical
congressional and legislative elections to evaluate four measures:
partisan bias, declination, efficiency gap, and mean-median
difference. While the efficiency gap recognizes simulated packing and
cracking in a completely predictable manner (a fact that follows
immediately from the efficiency gap's definition) and the declination
does a very good job of recording simulated packing and cracking, we
conclude that both of the other two measures record it poorly. This
deficiency is especially notable given the frequent use of such
measures in outlier analyses.
\end{abstract}

\maketitle 


\section{Introduction}

A gerrymander is a district plan that (dis)advantages a particular
group. Commonly, the group in question is a racial minority, a group
of incumbents or a political party. In this article we will focus on
cases in which the group is a political party --- partisan
gerrymanders. The clean separation between the various types of
gerrymanders given above is frequently not clear in practice. For
example, in \emph{Radogno \emph{v.}\ Illinois State Board of
  Elections},\footnote{Radogno v. Illinois State Board of Elections,
  NO. 1:11-cv-04884 (N.D. Ill. Oct. 21, 2011).} the 2011 legislative
district plan was challenged (unsuccessfully) as both a racial and a
partisan gerrymander. Similarly, in \emph{Cooper \emph{v.}\ Harris},
the 2011 North Carolina congressional plan was challenged as a racial
gerrymander;\footnote{Cooper \emph{v.}\ Harris, No. 15-1262 (S. Ct.).}
the resulting remedial plan was then challenged (unsuccessfully) on
remand as a partisan gerrymander.\footnote{Cooper \emph{v.}\ Harris,
  No. 16-166 (S. Ct.).} Regardless, we are still able to ask to what
extent a district plan qualifies as a partisan gerrymander, whether or
not it might qualify as a gerrymander of another type as well. For the
remainder of the article, unless otherwise noted, by ``gerrymander''
we mean ``partisan gerrymander.''

Historically, partisan gerrymanders have been recognized through their
contorted geographical boundaries. However, contorted shapes, while
often arising as a symptom of gerrymandering, are by no means a
necessary characteristic; contorted shapes may raise red flags, but
their absence is not particularly informative. Nonetheless, ways of
quantifying how contorted boundaries are --- ``compactness metrics''
--- continue to play a very important role in identifying
gerrymanders. They can be used to support allegations of intent by,
for example, showing that the challenged plan is significantly less
compact than expected based on a suite of computer-generated
plans. Perhaps more crucially, the creation of computer-generated
district plans requires some means of filtering out plans in which one
or more districts is sufficiently contorted, a purpose for which
compactness metrics are well suited.

The reason compactness measures are not a reliable means of detecting
gerrymanders is that gerrymanders are created by intentionally
allocating the voters of each party to the various districts in an
advantageous manner. The ``packing and cracking'' by which this is
done (discussed in more detail in Section~\ref{sec:how}) may require
contorted boundaries or it may not. Regardless, it is plausible that
one can identify gerrymanders not by looking at the shapes of
districts but rather by analyzing how the votes of each party are
distributed among the districts. A number of \emph{partisan-asymmetry
measures} have been proposed with this idea in mind. Unfortunately,
there is no consensus as to how to determine which of these measures
are successful. The main goals of this article are to
\begin{enumerate}
  \item Introduce a new technique, \emph{simulated packing and
    cracking (\spc)},\footnote{\ifdef{\SUBMIT}{Redacted.}{An early version of this technique
    applied to the declination only appears in our
    preprint~\citep{Warrington-Net}.}} for evaluating the ability of
    the partisan-asymmetry measures to detect gerrymandering and
  \item Apply \spc\ to four gerrymandering measures using a large
    collection of historical elections in order to evaluate their
    ability to record the primary signal of partisan gerrymandering.
\end{enumerate}

Comparisons and evaluations of partisan-asymmetry measures already
exist in the literature. We mention some recent
instances. In~\citep{MSII} the authors explore the extent to which
several measures adhere to the \emph{Efficiency Principle}, which is a
monotonicity principle relating partisan advantage, seats and
votes.\footnote{The efficiency principle is closely related to the
\spc\ Criterion introduced below; see~\citep{Veomett} for further
explorations of the efficiency principle in the context of the
efficiency gap.} In~\citep{McDonald-two}, the authors evaluate a
number of measures on two specific elections based primarily on their
stability under uniform vote swings. A similar approach to the prior
studies is taken in~\citep{Nagle1,Nagle2}, although with different
theoretical considerations and different measures considered. The
authors in~\citep{Campisi} focus on the ability of a specific measure
(declination) to deal with unequal turnout among
districts. In~\citep{compare}, the author considers how various
measures evaluate a number of hypothetical elections. Finally,
in~\citep{king}, the authors evaluate measures based on what they
contend are universal assumptions.

As mentioned above, there is essentially no agreement among the above
analyses as to which measures are the most effective. For some
situations, this appears not to matter. For instance, in
\citep{warshaw} the authors aim to quantify the political effects of
gerrymandering. For the four measures they consider --- the same four
we consider here --- there is general agreement as to the small but
measurable effects as averaged over a population of plans. Whatever
their strengths and weaknesses, each of the four measures comes to the
same general conclusion when considering plans en masse.  On the other
hand, there are many examples of specific elections for which specific
measures disagree strongly (see, e.g.,
\citep{compare,McDonald-two}). The existence of such discrepancies is
an important matter to consider more carefully.

The importance of characterizing the behavior and efficacy of
gerrymandering measures stems from their use in three different
contexts. First is their use in litigation. The possible utility of
quantitative social science in partisan-gerrymandering litigation was
crystallized in \emph{Davis \emph{v.}\ Bandemer},\footnote{478
U.S. 109 (1986)} which established that district plans could be
challenged as partisan gerrymanders. Quantitative tools were
incorporated in successive cases such as \emph{Whitford
\emph{v.}\ Gill},\footnote{\emph{Whitford \emph{v.}\ Gill},
No. 15-cv-421, F. Supp. 3d (2016).} and \emph{Rucho \emph{v.}\ League
of Women Voters}.\footnote{\emph{League of Women Voters of North
Carolina \emph{v.}\ Rucho},
No. 1:16-C\textsc{v}-1164-textsc{WO}-\textsc{JEP}, 2016.} Ultimately,
the US Supreme Court decided in \emph{Rucho \emph{v.}\ Common Cause}
that partisan gerrymandering claims are non-justiciable, thereby
closing off the federal courts to partisan gerrymandering
claims. However, we expect measures to continue to be used at the
state level, as recently seen in \emph{League of Women Voters of
Pennsylvania \emph{v.}\ Commonwealth of
Pennsylvania},\footnote{Petitioner's Opening Brief (Public Version),
No. 159 MM 2017, Jan. 15, 2018. Available
at~\url{https://www.brennancenter.org/sites/default/files/legal-work/LWV_v_PA_Petitioners-Brief.pdf}.}
and \emph{Common Cause \emph{v.}\ Lewis}.\footnote{Complaint, Wake
County Superior Court, No. 18-cv-014001,
~\url{https://www.brennancenter.org/sites/default/files/legal-work/Common-Cause-v.-Lewis-Complaint-FILED-Nov-13-2018\%20\%281\%29.pdf}.}
The bar for measures used in the legal context is potentially very
high as the ramifications of false positives/negatives may be
significant.

Second, partisan asymmetry measures can be used to flag
plans as potentially unfair (although one must expect false positives)
or fair (although one must expect false negatives). These measures
could be applied to maps proposed by a redistricting committee or to
maps proposed by citizens. Functionality in this vein is central to
the purpose of PlanScore \citep{planscore}. Such measures could be
useful in citizen-focused redistricting tools such as
Districtr~\citep{districtr} as well. In fact, a number of measures are
now available in Dave's Redistricting App~\citep{DRA}.

The third category is a catch-all for uses that are essentially
independent of the redistricting process. The media may want to use
compactness measures to illustrate in a memorable way to their readers
the districts with the most contorted boundaries (see, for example,
\citep{wp}). Or researchers may wish to study the evolution of
gerrymandering over the decades or to measure the impacts of
gerrymandering on political parties such as in \citep{warshaw}. The
characteristics needed of the measures will vary widely in this
category of uses.

The partisan-asymmetry measures we consider in this article all
associate a single number to a given election/district plan. The
simplest approach to using these numbers is to consider them in
absolute terms. For example,~\citep{M-S} proposed that in a legal test
for gerrymanders, a state would have to provide justification for any
plan with an efficiency gap of greater than 0.08 in absolute value. 

One way to provide context for the measure values is via computer
simulations. In several recent cases that have been litigated (for
example, ~\citep{pa-lwv}), plaintiffs have focused on how the score on
the plan of interest compares to the distribution of scores arising
from computer-generated plans. By incorporating the realities that
mapmakers have to take into account when drawing maps, such
simulations can provide support for, or undercut, lawmakers'
justifications provided for why the maps were drawn as they were. If,
say, the distribution of efficiency gap scores is a bell curve with 95
percent of the values between -0.12 and 0.12, then a score of 0.20 is
an ``extreme outlier''; one would expect to see such a large value
arise from a computer-generated plan only rarely. From this point of
view, the efficiency gap serves as a marker for putative gerrymanders
in the same way that contorted boundaries do. The logic of this
approach depends on the relative magnitude of the efficiency gap score
(or whatever measure is being used) being a good proxy for the
severity of the gerrymander. In other words, high values of measure
must be strongly correlated with severe gerrymanders. As we show in
Section~\ref{sec:analysis}, two of the measures we consider fail this
requirement to the extent that they fail to satisfy the\\

\textbf{\spc\ Criterion:} A partisan-gerrymandering measure should
reliably detect packing and cracking due to one or more applications
of \spc.\\

The organization of this article is as follows. In the next section we
introduce the four measures we are studying along with the
gerrymandering technique of packing and cracking. Simulated packing
and cracking, which consists simply in modifying the vote
distributions of an election in a manner consistent with how someone
aiming to create a gerrymandering would do so, is introduced in
Section \ref{sec:spc}. An analysis of the various partisan-asymmetry
measures appears in Section \ref{sec:analysis}. We close in
Sections~\ref{sec:disc} and~\ref{sec:conc} with a discussion and
summary of our results; we conclude, in part, that the mean-median
difference and partisan bias do a very poor job of recognizing the
signal of \spc.

\section{How can gerrymanders be quantified?}
\label{sec:how}

\subsection{Partisan-asymmetry measures studied}
\label{sec:measures}

We focus in this article on four measures that have been proposed to
quantify partisan gerrymandering:
\begin{enumerate}
  \item \emph{(Partisan) bias} \citep{GK}, \bias, compares the
    predicted share of seats a party would win if they won fifty
    percent of the statewide vote (determined using a uniform vote
    swing) to a share of fifty percent of the seats.

  \item \emph{Declination} \citep{declination}, \dec, is a measure of 
    differential responsiveness derived from the average winning
    margins of each party and the fractions of seats won.

  \item \emph{Efficiency gap} \citep{McGhee,M-S}, \eg, measures the
    relative number of votes ``wasted'' by each party. In this article
    we will assume there is equal turnout in each
    district. (In~\citep{mcghee:measure}, McGhee proposes a
    modification to his original definition that works well even if
    turnout is unequal.)

  \item \emph{Mean-median difference} \citep{McDonaldBest}, \mm, a
    standard technique for evaluating the asymmetry of a distribution,
    encodes the difference between the mean of the Democratic vote
    shares among districts and the corresponding median value. 

\end{enumerate}
Each measure associates a number to an \emph{$N$-district election},
defined as a weakly increasing sequence $\bl =
(\ell_1,\ell_2,\ldots,\ell_N)$ in which $\ell_i$ indicates the
Democratic fraction of the two-party (legislative) vote in district
$i$. Partisan bias and the efficiency gap associate a number between
-1/2 and 1/2 while the other two measures output a number between -1
and 1. It is important to note that the measures we consider
invariably output different values from election to election; the
``fairness'' of each district plan is filtered through the lens of
individual elections.

Our choice of measures to consider is guided by two
considerations. The efficiency gap, mean-median difference, and
partisan bias are the most widely used and discussed of the measures
that have been proposed. For example, these are the three measures
implemented by PlanScore \citep{planscore}. We include the declination
along with these three because\ifdef{\SUBMIT}{}{, as discussed in
  \citep{compare},} we believe it to be at least as efficacious, if
not more so, than these other three measures.\footnote{To our
knowledge, these are the only partisan-asymmetry measures that have
played a role in litigation, although for the declination it is only
through expert witness testimony in the federal district court cases
in Michigan (\emph{League of Women Voters of Michigan
\emph{v.}\ Benson}, Case 2:17-cv-14148-ELC-DPH-GJQ) and Ohio
(\emph{Ohio A. Philip Randolph Institute \emph{v.}\ Householder},
Case: 1:18-cv-00357-TSB-KNM-MHW, Plaintiff's Proposed Finding of Fact,
IV.A.5.); the others feature in multiple additional cases.}

Each of the measures listed above has either minor or major variants
that could be considered as well. The partisan bias measure considers
the seats allocated to each party at the 50\%-vote level on the
\emph{seats-votes} curve. One variant compares the seats that would be
won by each party at the observed statewide vote level $V$ as compared
to $1-V$ (see~\citep{king} for a more extensive description of these
interconnected notions). In~\citep{declination}, a minor variant of the
declination is introduced with the goal of attaining a muted response
in elections in which one party wins the vast majority of the
seats. The original version of the efficiency gap, defined in terms of
\emph{wasted votes} applies to elections in which turnout is allowed
to vary among districts. See~\citep{Veomett} for a discussion of some
of the ramifications of assuming equal turnout for the efficiency gap
as we do. See~\citep{Nagle1} for a discussion of the mean-median
difference and some related measures.

\subsection{Packing and cracking}

The ``packing and cracking'' used to allocate voters advantageously
works as follows. Suppose the Democrats are in control of
redistricting and the Republicans are poised to win district $X$. In
\emph{packing}, Republicans are moved from $X$ to other districts in
which the Republicans already have enough strength to win. These votes
are effectively wasted in the new districts while district $X$ falls
to the Democrats. \emph{Cracking} works similarly, except now the
Republicans are spread among districts that they have no chance of
winning. Once the cracking occurs, the recipient districts are lost by
the Republicans by smaller margins, but they are still
lost by them. 

\begin{figure}
  \centering
  \includegraphics[width=.8\linewidth]{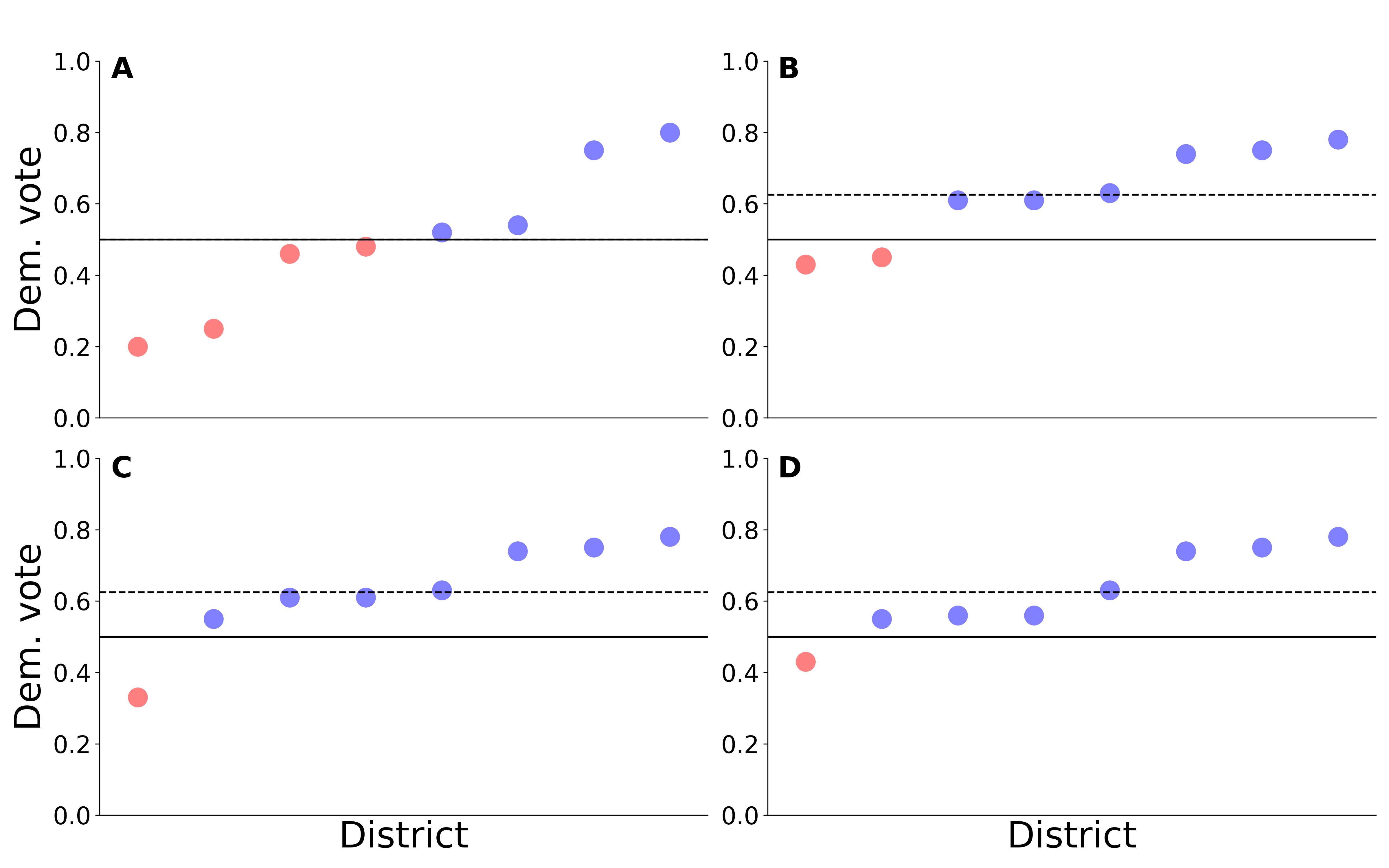}
  \caption{{\bf (A)} Symmetric, $8$-district election with four
    competitive districts. {\bf (B)} Hypothetical election deemed fair
    by all four measures (\bias, \eg\ and \mm\ all zero;
    \dec\ approximately $0.01$). {\bf (C)} Result of applying
    pro-Democratic packing \spc\ to election of A. {\bf (D)} Result of
    applying pro-Democratic cracking \spc\ to election of A. Statewide
    Democratic vote share in A, B and C cases is indicated by the
    dashed line at $0.625$.}
  \label{fig:pandc}
\end{figure}

In discussing packing and cracking, it is helpful to be able to
visualize how the voters of each party are distributed among
districts. We visualize the vector of district-level Democratic vote
shares, $\bl$, by plotting a point $v_i = (i/N-1/2N,\ell_i)$ for each
$i$. For example, Figure~\ref{fig:pandc}.A plots the symmetric vote
distribution $(0.2,0.25,0.46,0.48,0.52,0.54,0.75,0.8)$. In subplot B
we illustrate the vote distribution
$(0.43,0.45,0.61,0.61,0.63,0.74,0.75,0.78)$, for which the statewide
Democratic vote share is $0.625$. We have chosen the distribution so
that it is marked as fair by all four measures considered in this
article. Subplots C and D illustrate the effect of pro-Democratic
\pandc\ on the vote distribution illustrated in
Figure~\ref{fig:pandc}.B. Note that in Figure~\ref{fig:pandc}.C, the
second district has flipped to Democratic; Republican voters from the
second district were placed in the first district, thereby making a
single, strongly Republican district. In Figure~\ref{fig:pandc}.D, the
same district has been flipped, but this time Republican voters from
the second district have been redistributed to the third and fourth
districts.

While packing and cracking is the general technique by which
gerrymanders are created, in reality there will be ambiguities and
unknowables. For example, it is unreasonable to suppose that there is
a single, well-defined ``starting'' plan to which the packing and
cracking are applied. Certainly the map drawers may use the plan
currently in use as a starting point, but they may work from other
plans as well. Additionally, there is no reason to suppose that the
final map is drawn all in one step. A much more natural approach is to
iteratively progress towards a map that balances reward with the map
drawers' tolerance for risk and other priorities.\footnote{See, for
  example, the sequence of maps proposed for the North Carolina
  Superior Court,
  \url{https://www.ncleg.gov/Legislation/SupplementalDocs/2017/H717maps/H717maps}.}

\subsection{Simulated packing and cracking}
\label{sec:spc}

In order to validate the measures described in Section
\ref{sec:measures}, we will examine how perturbations of a vote
distribution by packing and/or cracking affect the value of each
individual measure. The technique we will use is that of
\emph{simulated packing and cracking (\spc)}. In
short, we manually modify a vote distribution by packing or cracking
so as to flip a single district from one party to the other. There are
four possible composite choices: whether we are packing or cracking
and whether it is the Republicans or the Democrats who are in charge
of the gerrymander. In reality, the votes from the flipped district
could be distributed among other districts by a combination of packing
and cracking, but we do not attempt to model this. We focus in the
below explanation on the case in which the Republicans are flipping a
single Democratic district to Republican control using cracking. The
other three cases are treated similarly.

In practice, the details of a gerrymander will depend on many
factors. One such factor will be the geography of the state. If a
given district is being cracked so as to turn it from a Democratic
district to a Republican district, the surplus Democratic voters will
have to be allocated to adjacent districts. We
make two observations. First, the geographic distribution of voters
may require additional swaps among more remote districts as we move
out from the flipped district, possibly affecting the entire district
plan. Second, the wave of modifications may require not only
significant deviations from the original boundaries, but also
extensive contortions of them.

As suggested in the previous paragraph, there are practical limits by
which actual gerrymanders are constrained. The \spc\ technique ignores
geography completely and focuses solely on what is numerically
possible. However, this does not mean that a plan resulting from a
geographic application of \spc\ are inherently infeasible. In
Section~\ref{sec:sup-geo} of the supplemental material, we illustrate
how one might apply \spc\ to the Indiana 116th congressional map while
still taking into account the geography of the state. The resulting
plan is markedly different from the starting one, and we make no claim
that the resulting boundaries are politically defensible, but the
shapes of the districts are no more tortured than many seen among
congressional districts plans from around the country.

Regardless, the measures we consider in this article should, if they
are effective measures, record packing and cracking regardless of
whether or not it geographically feasible. Said another way, while the
vote distributions we work with in this article are taken from
historical election data (and hence derive from underlying physical
district plans), mathematically the geography is irrelevant to the
questions we ask.

Another factor that affects the details of a gerrymander is how risk
averse the gerrymandering party is. For example, if the Democrats wish
to maximize their potential gain in seats (albeit at a high risk of
the plan backfiring) they can crack Republican districts by creating
districts that are (say) 49\% Republican. On the other hand, if the
Democrats feel the political winds will be against them in the
upcoming decade, they may prefer to pack Republicans into districts so
that the Democratic districts are no more than (say) 35\%
Republican. 

As stated above, \spc\ amounts to flipping one district by
redistributing some of the voters from the given district. In the
default case, we move some of the Democratic voters out from a
district so that the Republicans win that district without giving up
their hold on any other. To operationalize this idea, we make the
following conventions for how the gerrymander is achieved. It is
important to note that in some situations it will not be possible to
follow the conventions. In such situations, the attempted application
of \spc\ fails.
\begin{enumerate}
\item \emph{The district selected to be flipped is the Democratic district that
  is won by the narrowest margin.}\label{ci}
\item \emph{The gerrymander does not create any new
  Republican-majority districts with a Democratic vote fraction of
  greater than $0.45$.} We choose this value on the basis that a 45--55
  split is frequently considered the threshold for a race to be
  competitive (see, for example,~\citep{abramowitz}). Any
  Republican-majority district with a Democratic vote fraction higher
  than this before the cracking is allowed to remain at such a
  level.\label{cii}
\item \emph{The modified Democratic vote fraction in the flipped
  district is set to be $0.45$.}\label{ciii}
\item \emph{The Democratic votes shifted from the flipped district
  are distributed evenly among the Republican-majority districts
  with a Democratic vote fraction of at most $0.45$.} In order to
  avoid violating the Convention 2, this process may need to be
  iterated (see Example~\ref{lab:example} below).\label{civ}
\end{enumerate}

In order to illustrate the method in practice we present the following
example of flipping a district from Democratic to Republican by
cracking.  While we use hypothetical data in this example, all
subsequent applications of simulated packing and cracking in this
article begin with vote distributions from actual elections.

\begin{example}\label{lab:example}
  Consider a $10$-district election 
  \begin{equation*}
    \bl = (0.37,0.40,0.43,0.46,0.60,0.63,0.66,0.69,0.72,0.75).
  \end{equation*}
  By Convention~\ref{ci}, we flip the fifth district. By
  Convention~\ref{ciii}, the Democratic vote fraction in this district
  gets changes from $0.6$ to $0.45$. In order to maintain the same
  statewide Democratic vote fraction, there must be a net increase of
  $0.15$ among the first three districts (note that the fourth
  district is not included since its Democratic vote fraction already
  exceeds $0.45$). Convention~\ref{civ} instructs us to distribute
  these Democratic votes evenly among the three districts. The
  resulting vote distribution is
  \begin{equation*}
    (0.42,0.45,0.48,0.46,0.45,0.63,0.66,0.69,0.72,0.75).
  \end{equation*}
  However, following Convention~\ref{civ}, we iterate the process by
  redistributing the excess fraction of $0.03=0.48-0.45$ from the
  third district evenly among the first two districts. Since the
  second district is already at a value of $0.45$, the amount is
  entirely distributed to the first district. This yields a
  final\footnote{According to the definition of an election used in
  this article, the $\ell_i$ should be weakly increasing. We have not
  performed this final reordering step so as to aid the reader in
  following the redistribution of votes.} vote distribution of
  \begin{equation*}
    \bnl = (0.45,0.45,0.45,0.46,0.45,0.63,0.66,0.69,0.72,0.75).
  \end{equation*}
\end{example}

In Section~\ref{sec:sup-pack} of the supplemental material we perform an analogous computation for a
pro-Republican, packing application of \spc.

\section{Evaluations of measures using \spc\ applied to historical elections}
\label{sec:analysis}

We now examine how the values of the measures introduced in
Section~\ref{sec:measures} change in response to simulated
packing and cracking.

The starting vote distributions, to which \spc\ will be applied, are
taken from two different collections of historical election data, both
described in detail
in \ifdef{\SUBMIT}{(redacted)}{~\citep{declination}}. The first
collection consists of congressional elections since 1972 while the
second collection consists of elections to state lower-house
legislatures since 1972.\footnote{A multilevel model (described
in\ifdef{\SUBMIT}{(redacted)}{~\citep{declination}}) is used to impute
the Democratic vote fraction for uncontested elections.} For each such
election we attempt the four possible types of \spc: packing in favor
of the Democrats, packing in the favor of the Republicans, cracking in
favor of the Democrats and cracking in favor of the Republicans. As
noted above, our applications of \spc\ do not take into account any of
the geographic aspects of the district plans from which our historical
election data are taken. In light of this, our investigations shed no
light on the historical elections themselves.

\begin{figure}
  \centering
  \includegraphics[width=1\linewidth]{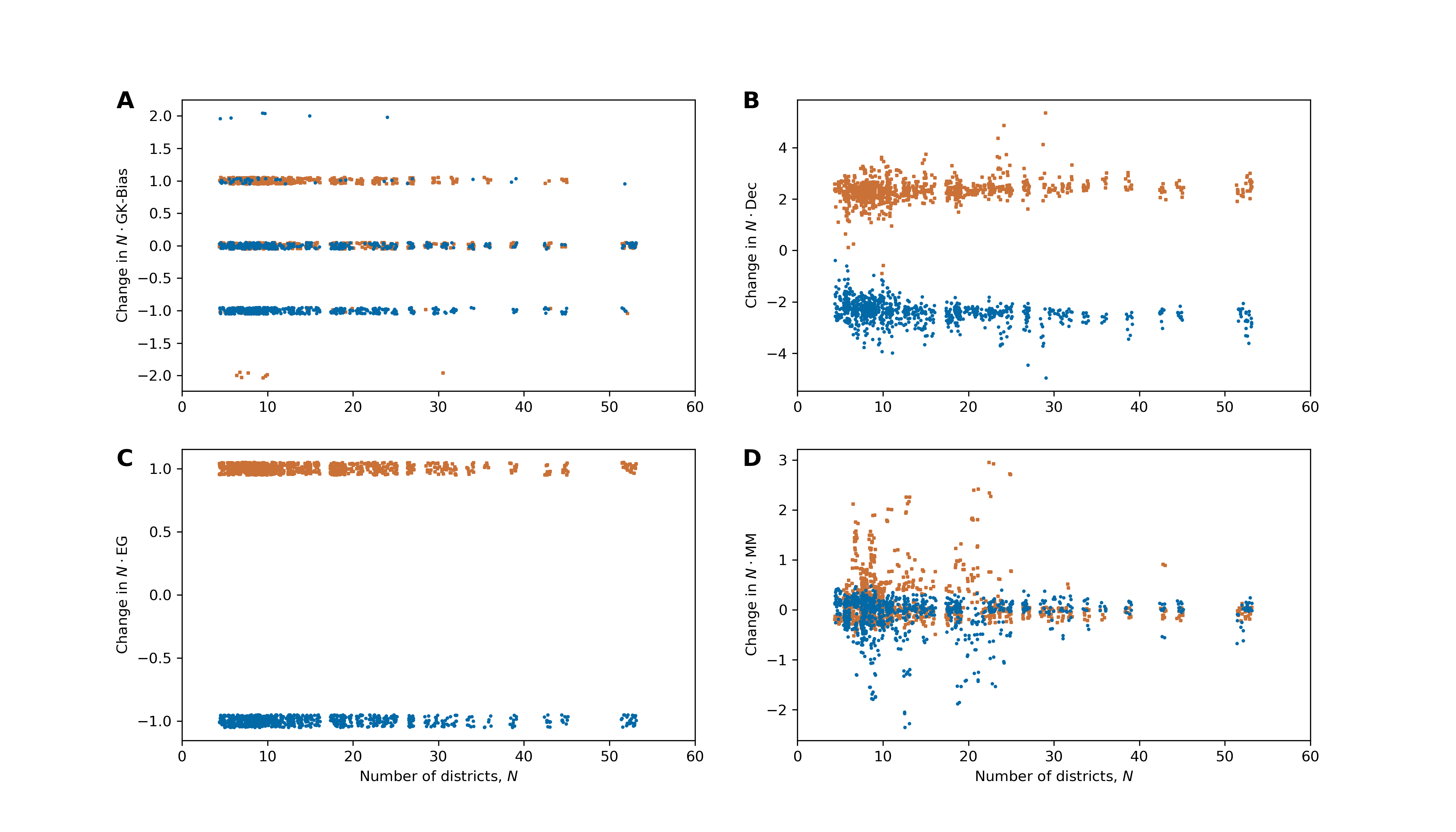}
  \caption{Plot of change in measures (each scaled by $N$) due to
    \spc. For each subplot, the 961 red squares represent single
    applications of pro-Republican \spc\ to House elections since
    1972; the 939 blue circles represent single applications of
    pro-Democratic \spc.
    Measures depicted are partisan bias (A), declination (B),
    efficiency gap (C) and mean-median difference (D). Horizontal and
    vertical coordinates have been jittered to reduce overlap.
}
  \label{fig:pandc-1-mpandc-11}
\end{figure}

Our first exploration is of how the value of each measure changes
under a single application of SPC. Ideally, flipping a single seat
from (say) Democratic to Republican would lead to a consistent
increase in the measure value, regardless of the starting
distribution. Empirically, however, there is an inverse dependency on
the number of districts $N$. 

In fact, it follows from the definition of the efficiency gap that
flipping a single seat for an $N$-district election will lead to a
change of exactly $1/N$ in its value. While the partisan bias need not
change by exactly $1/N$, it will change by a multiple of
$1/N$. Empirical data \ifdef{\SUBMIT}{(see
  [redacted])}{(see~\citep{Warrington-Net})} as well as the figures in
this article) suggest that the declination changes by approximately
$1/2N$. Given these relationships, we have chosen to focus on the
change in value of each measure \emph{multiplied by the number of
districts, $N$}. This exactly removes the dependency on $N$ for the
efficiency gap and partisan bias and appears to approximately remove
it for the declination.

Figure~\ref{fig:pandc-1-mpandc-11} displays four plots, one for each
of the measures we consider. Each point in each plot corresponds to a
single successful (see below) application of \spc\ to one of the
elections in our data set. Red squares denote pro-Republican
packing/cracking while blue circles denote pro-Democratic
packing/cracking. Packing versus cracking are not distinguished in
Figure~\ref{fig:pandc-1-mpandc-11}. The horizontal coordinate of each
\spc\ application is given by the number of districts $N$ in the
underlying plan to which \spc\ is being applied; the vertical
coordinate is the change in measure, scaled by $N$.

Table~\ref{tab:fits} contains details about the data of
Figure~\ref{fig:pandc-1-mpandc-11} as well as for the analogous figure
(not shown) for legislative elections. In Section~\ref{sec:sup-noise} of the supplemental material we include
material analogous to Figure~\ref{fig:pandc-1-mpandc-11} and
Table~\ref{tab:fits} but with statistical noise added to the vote
distributions both before and after the application of \spc.

\begin{table}
  \centering
  \caption{Success of each measure at moving the expected direction
    under one application of \spc\ for House and state-legislative
    elections.}
\begin{tabular}{@{}llcccccc@{}}\toprule
  & {} & & \multicolumn{2}{c}{pro-Democratic} & \phantom{ab} & \multicolumn{2}{c}{pro-Republican} \\ \cmidrule{4-5} \cmidrule{7-8}
  & & Fraction correct & Decrease & Increase & & Decrease & Increase \\\midrule
  & \bias & 0.531 & 508 & 431 & & 460 & 501 \\
  & \dec  & 0.999 & 939 &   0 & &   2 & 959 \\\rot{\rlap{House}}
  & \eg   & 1.000 & 939 &   0 & &   0 & 961 \\ 
  & \mm   & 0.455 & 422 & 517 & & 518 & 443 \\\midrule
  & \bias & 0.340 & 538 & 752 & & 951 & 338\\
  & \dec  & 1.000 & 1290 &   0 & &   0 & 1289\\
  & \eg   & 1.000 & 1290 &   0 & &   0 & 1289\\ \rot{\rlap{Legislative}}
  & \mm   & 0.308 & 428 & 862 & & 923 & 366\\\bottomrule
\end{tabular}
\label{tab:fits}
\end{table}

The instances of \spc\ plotted in Figure~\ref{fig:pandc-1-mpandc-11}
and the subsequent figures are restricted in two ways. First, we
require that each party win at least one seat both before and after
the packing/cracking. This is necessary for the declination of each
distribution to be defined. Second, we require that there be at least
three districts into which to distribute the votes from the flipped
district for each application of \spc. For example, when a seat is
being flipped from Democratic to Republican by cracking, we require
that there be at least three Republican seats in the original
distribution. Together, these restrictions exclude all elections from
states in which there are four or fewer congressional districts.\footnote{In
the current apportionment cycle there are 21 such states:
  AK, AR, DE, HI, IA, ID, KS, ME, MS, MT, ND, NE,
  NH, NM, NV, RI, SD, UT, VT, WV, and WY.}
There were 676 state-year pairs in which there were at least five
Congressional districts and each party won at least one seat. Given
the four possible combinations of packing/cracking and
pro-Republican/pro-Democratic, this offers 2704 possible applications
of \spc. However, for 804 of these, either there was not enough room
for the chosen packing/cracking or one of the constraints was not
satisfied. Remaining are the 1900 simulations illustrated in
each subplot of Figure~\ref{fig:pandc-1-mpandc-11}.

\begin{figure}
  \centering
  \includegraphics[width=1\linewidth]{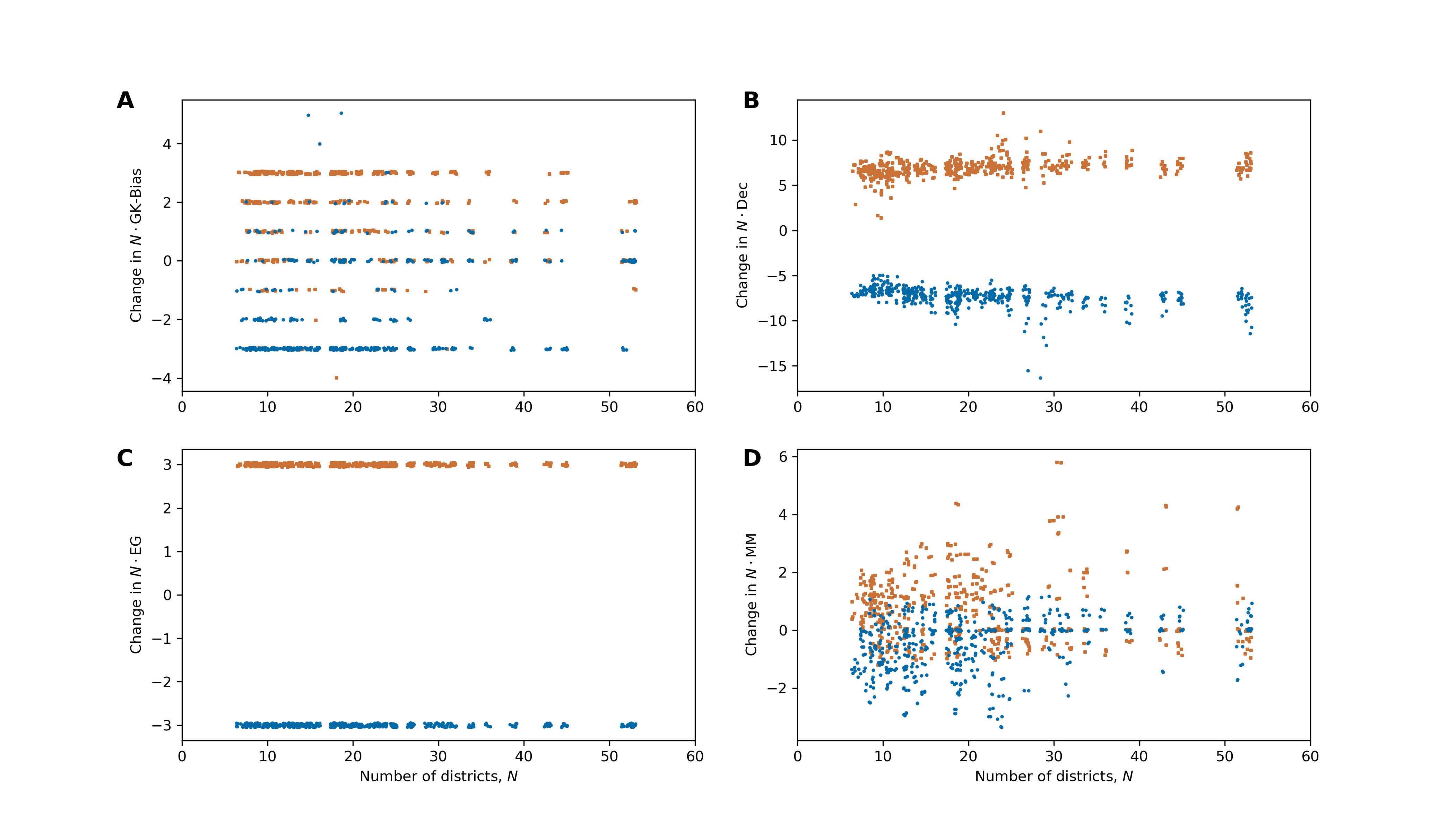}
  \caption{Plot of net change in measures (each scaled by $N$) due to
    three applications of \spc. For each subplot, the 540 red squares
    represent triple applications of pro-Republican \spc\ to House
    elections since 1972; the 529 blue circles represent single
    applications of pro-Democratic \spc.
    Measures depicted are partisan
    bias (A), declination (B), efficiency gap (C) and mean-median
    difference (D). Horizontal and vertical coordinates have
    been jittered to reduce overlap.
}
  \label{fig:pandc-3-mpandc-11}
\end{figure}

\begin{figure}
  \centering
  \includegraphics[width=1\linewidth]{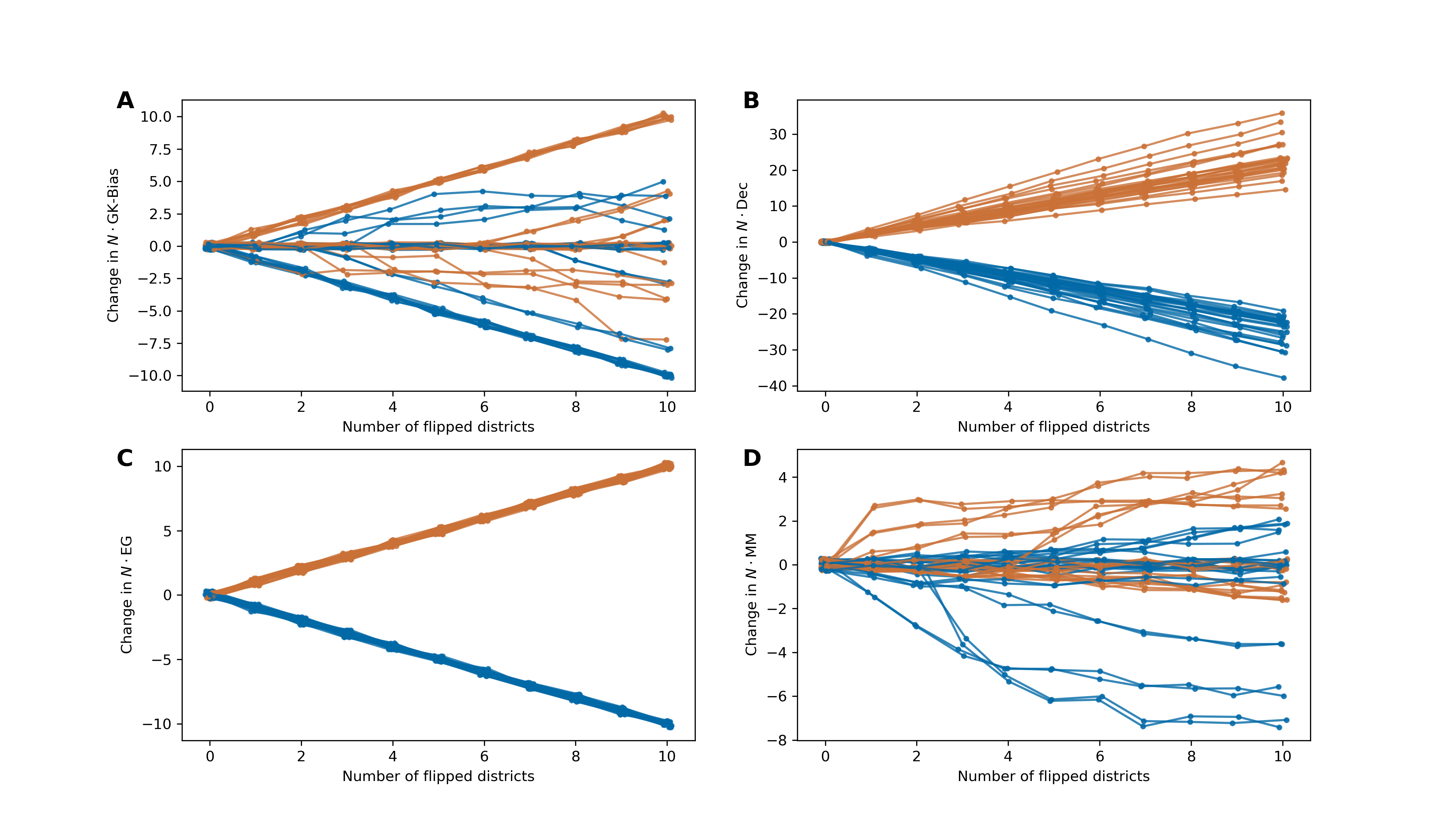}
  \caption{Plot of incremental change in measures under successive
    applications of \spc. Measures depicted are partisan bias (A),
    declination (B), efficiency gap (C) and mean-median difference
    (D).
  }
  \label{fig:pandc-seqs-10-9}
\end{figure}

Our second exploration is of how the measures change under the
flipping of multiple seats. Figure~\ref{fig:pandc-3-mpandc-11}
considers the net change after three such flips.\footnote{We have
  chosen three for the congressional elections as representative of
  what might happen in a severe gerrymander such as in North
  Carolina.} 
Axes are as in Figure~\ref{fig:pandc-1-mpandc-11}.
In Figure~\ref{fig:pandc-seqs-10-9}, we consider
the incremental effects. For example, in
Figure~\ref{fig:pandc-seqs-10-9}.B we show one line plot for each
legislative election considered.\footnote{We have only shown the line
  plots for a sample of 12 legislative elections (about 2\% of the
  data set) in order to avoid clutter.} For each election, we
successively apply \spc, evaluating the resulting vote distribution
under the declination after each application. Each election results in
up to four line plots, one for each of the four possible combinations
of packing/cracking and pro-Republican/pro-Democratic. In subplot B,
the pro-Republican line plots are increasing and the pro-Democratic
ones are decreasing. This indicates that larger changes in the
declination generally correspond to greater numbers of flipped seats
under \spc.

Our third exploration looks at the extent to which our results depend
on the implementation choices we have made for \spc. Two fundamental
choices we make are how close to parity we allow a modified district
to get and how votes from the flipped district are reallocated to the
receiving districts. Below we list the alternatives we consider
(phrased relative to a pro-Republican \spc).
\begin{enumerate}
\item Maximum competitiveness: The maximum Democratic vote share
  allowed in a modified district.
  \begin{enumerate}
  \item \emph{0.45:} Default.
  \item \emph{0.40:} Increased risk aversion relative to default.
  \item \emph{0.49:} Decreased risk aversion relative to default.
  \end{enumerate}
\item Algorithm: How voters from the flipped district are reallocated
  to the receiving districts.
  \begin{enumerate}
    \item \emph{Even:} Default. Distribute new Democratic voters
      evenly among the receiving districts, iterating as necessary.
    \item \emph{Equalization:} Distribute new Democratic voters
      preferentially to the \emph{least} Democratic districts. This
      has the effect of equalizing Democratic support among the
      Republican districts (when cracking) or among Democratic districts
      (when packing).
    \item \emph{Concentration:} Distribute new Democratic voters
      preferentially to the \emph{most} Democratic districts. This has
      the effect of accentuating differences in Democratic support
      among the Republican districts (when cracking) or among Democratic
      districts (when packing).
  \end{enumerate}
\end{enumerate}

\begin{figure}
  \centering
  \includegraphics[width=1.0\linewidth]{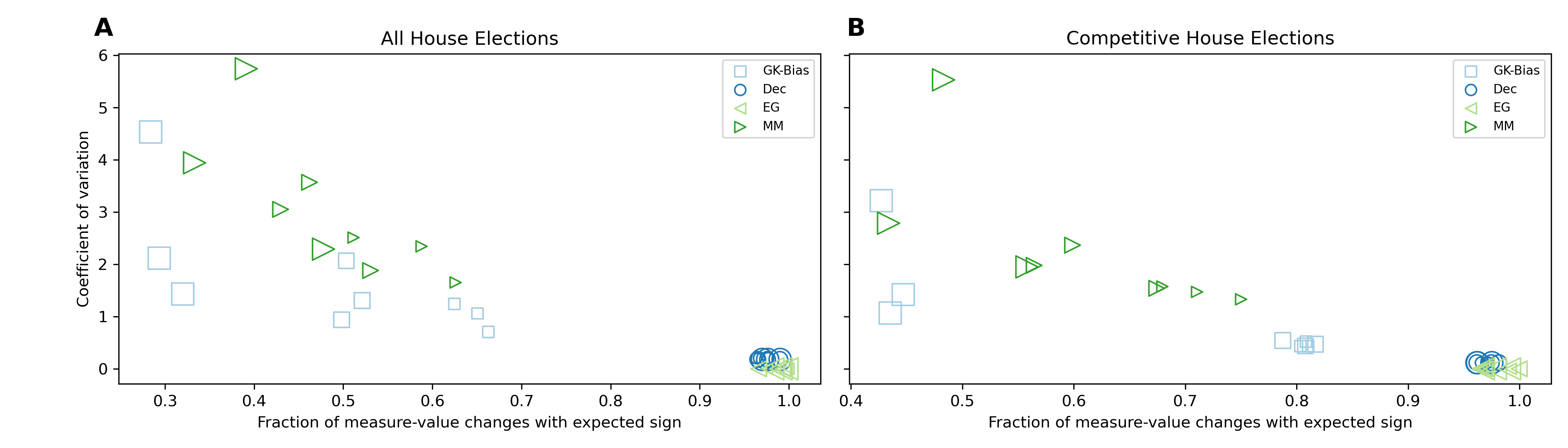}
  \caption{Fraction of elections for which each measure moves in the
    expected direction plotted against the uniformity of the change
    (as measured by the coefficient of variation) for each variation
    of the standard \spc\ algorithm. The sizes of markers denote the
    corresponding maximum competitiveness as described in the text;
    the redistribution algorithm used is not displayed
    graphically. Subplot \textbf{(A)} is based on all House elections
    in our data set, subplot \textbf{(B)} on those ``competitive''
    ones for which the statewide Democratic votes was between 45\% and
    55\%.}
  \label{fig:variations}
\end{figure}

When flipping a single seat, each of the nine combinations of choices
generates data analogous to that illustrated in
Figure~\ref{fig:pandc-1-mpandc-11}. As the numerous individual
subplots would be difficult to compare even qualitatively, we
summarize the results as a scatter plot in
Figure~\ref{fig:variations}.A. This figure contains 36 data points,
one for each of the nine \spc\ variations paired with each of the four
measures we consider.\footnote{We are not distinguishing in this figure
  between pro-Republican/pro-Democratic or packing/cracking} The
horizontal axis records the fraction of time the measure changes in
the expected direction while the vertical axis records the variation
in the change, appropriately scaled. Figure~\ref{fig:variations}.B
contains analogous data restricted to elections where each party enjoys
at least 45\% of the vote. We now describe in more detail how these
summary statistics are computed.

To determine the horizontal coordinate for a given \spc-measure pair,
we determine the fraction of successful applications of that variation
of \spc\ that move the associated measure in the expected direction:
Under a pro-Republican packing/cracking, we expect a given measure
value to increase in value (i.e., become more positive if initially
positive and less negative if initially negative). For example, for
the default \spc\ variation used for
Figure~\ref{fig:pandc-1-mpandc-11}.C, the efficiency gap increases for
every single pro-Republican packing/cracking and decreases for every
single pro-Democratic packing/cracking. This pair has a horizontal
coordinate of $1$ in Figure~\ref{fig:variations}. For the declination,
the fact that in Figure~\ref{fig:pandc-1-mpandc-11}.B, 959 out of 961
of the red squares lie strictly above $0$ while all 939 of the blue
circles lie strictly below $0$, yields a horizontal coordinate of
0.999.

The vertical coordinate quantifies the consistency of each measure in
recording packing and cracking, as measured by the coefﬁcient of
variation, defined as the sample standard deviation of the measure
divided by its sample mean.\footnote{In these computations, we
  multiply the differences for all the pro-Democratic applications of
  \spc\ by $-1$ so that the expected differences are always positive.}
Dividing by the sample mean accounts for the fact that the scalings of
the original measures are essentially arbitrary; this normalization
allows us to compare them on an equal footing. As an example, consider
again the declination as shown in
Figure~\ref{fig:pandc-1-mpandc-11}.B. The standard deviation of the
1900 data points is 0.428 while the corresponding mean is 2.36. The
ratio of 0.428/2.36 yields a coefficient of variation of 0.181.

In Figure~\ref{fig:variations}, color and shape denote measure as
indicated in legend. Small, medium and large markers denote maximum
competitiveness of 0.40, 0.45 and 0.49, respectively. The
redistribution algorithm used is not indicated graphically. 

\section{Discussion}
\label{sec:disc}


Our starting point for this article is that any effective
partisan-asymmetry measure should be able to consistently recognize
packing and cracking, the accepted means by which
partisan-gerrymanders are created.\footnote{For example, ``For packing
  and cracking are the ways in which a partisan gerrymander dilutes
  votes,'' \emph{Gill \emph{v.}\ Whitford}, concurring opinion
  (J. Kagan),~\url{https://www.law.cornell.edu/supct/pdf/16-1161.pdf},
  page 4.} As a proxy for recognizing packing and cracking in
historical election data, which can be difficult to separate from
social and political factors, we have proposed the
\spc\ Criterion. Below we explain our understanding of why the
measures behave as shown in this article under applications of
\spc. For clarity of exposition, we describe the behavior in the
context of pro-Republican applications of \spc.

\emph{Efficiency gap:} As illustrated in Figures~\ref{fig:pandc-1-mpandc-11},
\ref{fig:pandc-3-mpandc-11} and~\ref{fig:pandc-seqs-10-9}, the
efficiency gap captures each seat flip with a change of exactly
$1/N$. It is important to note here that this depends in part on our
assumption of equal turnout in each district: As shown in
~\citep{McGhee}, if $V$ denotes the statewide Democratic vote share and
$S$ denotes the fraction of Democratic seats, then the value of the
efficiency gap reduces to $(S-1/2)-2(V-1/2)$. Since \spc\ does not
change $V$, flipping a seat (i.e., changing $S$ by $\pm 1/N$) changes the
efficiency gap in a completely predictable manner. This is arguably
the ideal outcome. If turnout is not assumed to be equal and one
computes the efficiency gap according to the more general notion of
wasted votes (see~\citep{McGhee}), its value will likely change by
amounts close to $1/N$, but not exactly equal to it.

Because of the strict relationship among seats, votes and the value of
the (equal-turnout) efficiency gap, \spc\ does not actually provide any
new insights in this case. However, we have included the efficiency
gap in this article for two reasons. First, it has been the most widely
used measure over the past several years. Second, its behavior
illustrates any arguably ideal way a measure should behave under
\spc. It therefore serves as a useful benchmark for the evaluating the
performance of the other measures considered.

The efficiency gap measure has come under strong criticism. This
criticism has run the gamut from its behavior when one party earns
more than 75\% of the statewide vote~\citep{MoonMira}, to its behavior
when there only a few districts~\citep{cho-upenn}, to its choice of how
to weight the various classes of ``wasted'' votes~\citep{Cover}. Some
of these criticisms (such as the behavior beyond 75\%) are well
founded but only serve to limit the situations in which the efficiency
gap can provide meaningful information. Others, such as the debate on
how to weight wasted votes, are central to the measure: The choice of
how to weight wasted votes directly determines what fraction of seats
a given party should (normatively) earn for a given statewide vote
share. The arguments for and against different choices are many and
there does not appear to be a single correct answer. One benefit of
simulated packing and cracking is that it allows us to explore the
extent to which measures record seats flipped by packing and
cracking. That is, \spc\ is concerned only with the \emph{change} in a
measure's value rather than its absolute value for any given
election. As discussed above, the efficiency gap records these changes
faithfully and we see this as a very positive aspect of the measure.

\emph{Declination: } As proved in Theorem~1 of~\citep{declination},
the declination will increase under a pro-Republican application of
\spc\ as long as the modified Democratic vote fraction in the flipped
district is greater than or equal to the average Democratic vote
fraction in the other Republican districts. That this is the case
follows from the geometric definition of the declination as an angle
along with three facts. First, that we are flipping the most
competitive (Democratic) district.\footnote{If we flip the second-most
  competitive district, the fraction of House elections for which the
  declination moves in the expected direction decreases from 0.999
  (see Table~\ref{tab:fits}) to 0.989.} Second, that the average Democratic
vote share will stay the same or increase in both the Democratic
districts and the Republican districts.\footnote{This is true
  regardless of whether voters in the flipped district are
  redistributed through packing or through cracking.} And third, that
the Republicans gain one seat.

The declination does a very good job of responding to \spc. In
Figures~\ref{fig:pandc-1-mpandc-11} and~\ref{fig:pandc-3-mpandc-11} it
shows a almost entirely clean separation between positive values for
pro-Republican packing/cracking and negative values for pro-Democratic
packing/cracking (note the two negative values for pro-Republican
packing/cracking in Figure~\ref{fig:pandc-1-mpandc-11}.B). And when
change in declination is considered incrementally as in
Figure~\ref{fig:pandc-seqs-10-9}, we see that it increases in a
regular manner for the sample shown,\footnote{This is true
  qualitatively for the entire data set as well.}  although at
slightly different rates for different individual elections.

Having discussed both declination and efficiency gap individually, it
is useful to contrast their behavior. The efficiency gap changes by a
fixed, predictable amount for every seat flip. The declination changes
consistently, but with some variation. Delving into the pros and cons
of each measure is beyond the scope of this article. However, we do
wish to point out that an argument can be made that a measure
\emph{should} change by varying amounts depending on the particulars
of how the vote distribution has changed. As just one example,
consider two applications of pro-Republican \spc\ that differ only in
whether the redistributed voters are packed or cracked. Under
cracking, these voters would be redistributed to Republican districts,
thereby making these districts slightly more Democratic. If these
districts are already reasonably competitive, the addition of
Democratic voters might tip them over into being highly
competitive. In such a scenario, it might be better for the
Republicans to redistribute these voters through packing to districts
that are already safely Democratic. The latter case could lead to a
``safer'' gerrymander that could reasonably be measured as being more
advantageous to the Republicans than the former version.

\emph{Partisan bias:} The partisan bias measure compares the seat
share each party would earn were the statewide vote shifted to
50\%. Suppose, as above that the statewide Democratic vote share is
$V$. The value of partisan bias is then determined by the fraction of
districts with a Democratic vote share of less than $V$. When \spc\ is
applied, the change in partisan bias is therefore determined by the fraction of
districts whose Democratic vote share switches from one side of $V$ to
the other. If the state is evenly split, the flipped district will
typically flip from above $V$ to below, increasing the partisan bias by
$1/N$. The increased likelihood of this with decreased values of the
maximum competitiveness can be seen in
Figure~\ref{fig:variations}.A. In this figure, the sizes of the
squares indicate the final Democratic vote share in the flipped
district. Each of the large, medium and small squares are cleanly
separated from the other two groups. Furthermore, the smaller the
square (and hence the greater the swing in the flipped-district
Democratic vote share) the more likely the partisan bias increases.

\begin{figure}
  \centering
  \includegraphics[width=.8\linewidth]{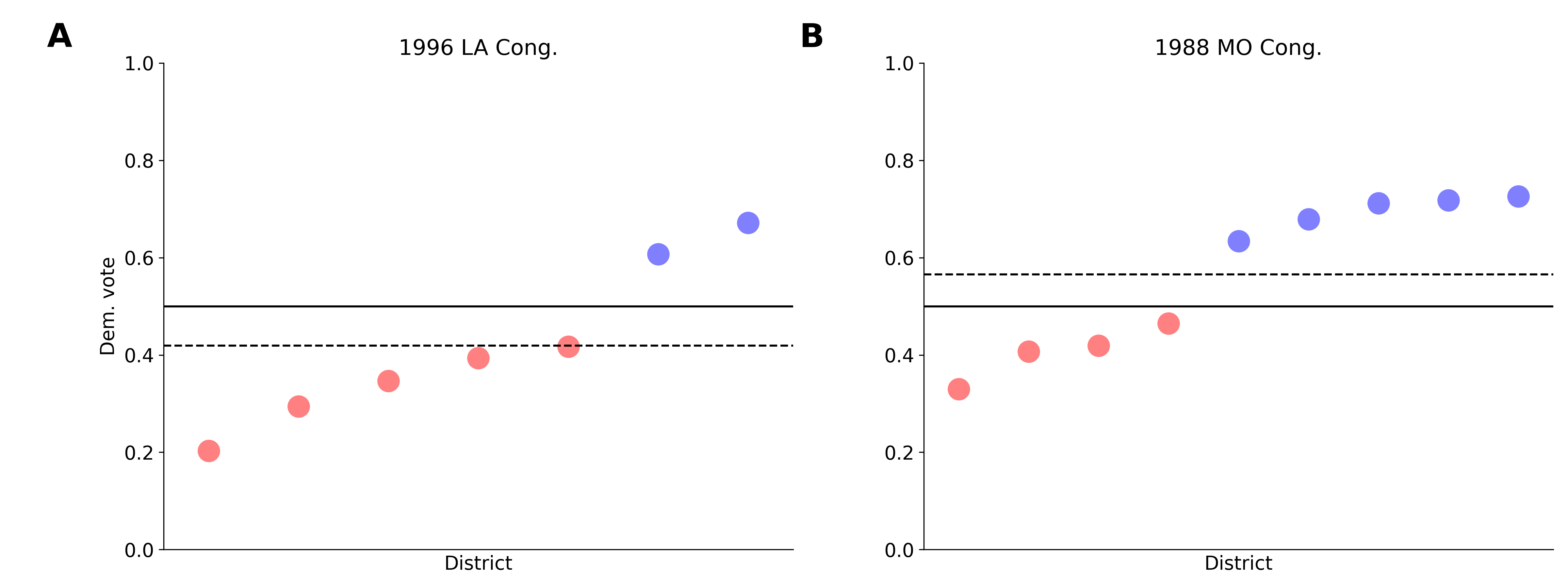}
  \caption{{\bf (A)} 1996 Louisiana House election (values for several
    uncontested races are imputed). {\bf (B)} 1988 Missouri House
    election.}
  \label{fig:disc}
\end{figure}

The redistribution of votes can cause other districts to move back
across $V$ in the opposite direction. This accounts for the
applications of \spc\ illustrated in
Figure~\ref{fig:pandc-1-mpandc-11}.A that move in the wrong
direction. As a concrete example consider a seven-district election
with Democratic votes shares of 0.203, 0.294, 0.347, 0.394, 0.417,
0.607 and 0.672, as depicted in
Figure~\ref{fig:disc}.A.\footnote{These values come from the 1996
Louisiana House election, although several values are imputed.} The
statewide average is 0.419. When the statewide average is shifted to
0.5, the number of Republican districts remains at five. Under a
pro-Republican, cracking application of \spc, the original district
values become 0.234, 0.325, 0.378, 0.425, 0.448, 0.450 and 0.672. As
the statewide average doesn't change, only three Republican districts
remain when the statewide average is shifted to 0.5, leading to a
change in partisan bias of $-2/7$, even though the change should be
positive as it is derived from a pro-Republican application of \spc.

\emph{Mean-median difference: } When \spc\ is applied to a vote
distribution, the statewide Democratic vote share does not change. Any
change to the measure therefore results from a new value for the
Democratic vote share in the median district. While its value often
increases as expected, as seen in Figure~\ref{fig:pandc-1-mpandc-11}.D
it frequently \emph{decreases}, though only by small amounts. This can
happen when some of the Democratic voters moved out of the flipped
district are redistributed to the median district. The same figure
also shows high variability when it does move in the expected
direction. 

The 1988 Missouri House election shown in Figure~\ref{fig:disc}.B is
useful for highlighting some of the tendencies of the measures we have
been discussing in the abstract. The vote distribution is reminiscent
of a bipartisan (i.e., incumbent-protection) gerrymander, with at most
one competitive district. Suppose we apply \spc\ via pro-Democratic
packing, flipping the fourth district from Republican to
Democratic. As a packing gerrymander, the Democratic vote share in the
median district remains unchanged; the mean-median difference is
insensitive to this application. The partisan bias is also unable to
register this flip as the new Democratic vote share in the flipped
district of 0.55 is less than the statewide Democratic vote share of
0.566. Both the declination and the efficiency gap, by incorporating
the number of seats won by each party into their values, do register
the flip.

Figure~\ref{fig:variations} depicts two additional facets of measures
under \spc. First, the vertical coordinate is a measure of how uniform
the change to a given measure is under \spc\ as the underlying
election varies. In general, lower coefficients of variation are
preferable given that we are attempting to analyze district plans
rather than elections per se. Second this figure summarizes how
dependent the results for each measure are on the structure of the
\spc\ model. As shown by the tight clustering for the declination and
efficiency gap,\footnote{Markers are jittered to reduce overlap.}
these two measures are insensitive to the implementation details of
\spc. The clouds for partisan bias and mean-median difference show
there is a reasonable amount of dependency, yet these two measures do
not do particularly well for any of the \spc variations.

One commonality between the efficiency gap and the declination is that
they are averaging vote shares over districts. This explains the
stability for these measures illustrated in
Figure~\ref{fig:variations} under varying the redistribution
algorithm. These measures are, of course, still sensitive to other
features such as the fraction of seats won by each party. Partisan
bias and the mean-median difference are more sensitive to what happens
in specific districts; this feature appears to be responsible at least
in part to the lower success rates of these measures. As explored in
Section~\ref{sec:sup-noise} of the supplemental material, incorporating uncertainty into our district-level vote
shares changes the calculus somewhat; the declination and efficiency
gap do not behave as predictably. While they are consistently more
successful in recording \spc\ for the House elections, there is little
to choose between the four measures in the legislative elections,
likely due to the frequently large number of competitive districts in
the large legislative plans.

As mentioned in the introduction, a number of analyses of
gerrymandering measures exist in the literature. We briefly discuss
one of the most recent and prominent studies as its conclusions differ
markedly from our own. In~\citep{king}, the authors purport to explore
the ``[t]heoretical foundations\ldots of partisan fairness,'' and
conclude that partisan bias is the only measure satisfying various
nonnegotiable requirements. But implicit to their approach is the
conflation of partisan fairness with symmetry of the seats-votes
curve. While symmetry of this curve is a reasonable desideratum, it is
by no means necessary. By taking it is a necessity, the authors give
equal importance to the counterfactual election results corresponding
to all possible uniform vote swings. So, for example, in a state that
reliably has 42\% statewide Democratic support, any evaluation of the
district plan gives equal importance to the hypothetical results you'd
get by modeling what would happen under 50\%, or even 58\%, statewide
Democratic support. This framework has a number of appealing aspects,
but, as our results suggest here, it does not appear to capture
packing and cracking. Regardless, King et al. are focused on what
happens under a range of Democratic vote shares while \spc\ leaves the
statewide Democratic vote share constant.

While we find no evidence here that partisan bias and the mean-median
difference are good measures of partisan gerrymandering, they still
are able to get at something very important: How the situation in
which each party wins half the seats relates to the situation in which
each party wins half the votes (see,
e.g.,~\citep{mcghee:measure}). There is no question that, in the
abstract, these two situations should coincide in a fair
election. Each of partisan bias and the mean-median difference measure
the degree to which these notions don't align. This is useful
information, but it is distinct from a quantitative measure of
partisan gerrymandering.

\section{Conclusion}
\label{sec:conc}

We have introduced simulated packing and cracking as a technique for
examining how partisan-gerrymandering measures respond to the type of
vote-distribution modifications that occur in partisan
gerrymandering. Partisan bias and mean-median difference are unable to
consistently record simulated packing and cracking, even when
restricted to competitive elections while the declination and (by
definition) the efficiency gap do well. As a result, we recommend that
neither partisan bias nor the mean-median difference be used for the
``outlier'' or ``ensemble'' method, where it is crucial that more
extreme values of the measure indicate more extreme levels of partisan
gerrymandering.

\section{Acknowledgments}

\ifdef{\SUBMIT}{Redacted}{
The US congressional data through 2014 was provided
by~\citep{jacobson}. The election data was analyzed using the
python-based SageMath~\citep{sage}. See~\citep{declination} for packages
used for, and details of, the imputation of votes. Additional Python
packages employed in this article were Matplotlib~\citep{matplotlib} for
plotting and visualization and SciPy~\citep{scipy} for statistical
methods. \ifdef{\SUBMIT}{}{All non-library code may be found at~\citep{tarball}.}
We thank Jordan Ellenberg for suggesting one of the \spc\ algorithms
for~\citep{Warrington-Net} and to Eric McGhee, John Nagle and Ellen
Veomett for helpful comments on a draft of this article.}

\bibliography{gerrymandering}

\bibliographystyle{apa-good}

\clearpage
\section{Supplemental Material}

\vspace*{.2in}
\begin{center}
  Appendix to\\
  \textbf{Simulated Packing and Cracking}\\
  \ifdef{\SUBMIT}{}{by Jeffrey S. Buzas and Gregory S. Warrington}
\end{center}

\vspace*{.2in}

\subsection{Packing example}
\label{sec:sup-pack}

We complement the cracking \spc\ example from the main text with an
example illustrating pro-Republican packing.

\begin{example}\label{lab:example2}
  Consider the same $10$-district election as from the main article:
  \begin{equation*}
    \bl = (0.37,0.40,0.43,0.46,0.60,0.63,0.66,0.69,0.72,0.75).
  \end{equation*}
  By Convention~\ref{ci}, we flip the fifth district. By
  Convention~\ref{ciii}, the Democratic vote fraction in this district
  gets changes from $0.6$ to $0.45$. In order to maintain the same
  statewide Democratic vote fraction, there must be a net increase of
  $0.15$ among the last five districts. We now redistribute these
  voters according to the three algorithms considered:
  \begin{itemize}
    \item \emph{Even:} Each of the five districts increases by $0.03$,
      leading to a final vote distribution of
      \begin{equation*}
        (0.37,0.40,0.43,0.46,0.45,0.66,0.69,0.72,0.75,0.78).
      \end{equation*}
    \item \emph{Equalization:} We first increase the sixth district
      from $0.66$ to $0.69$ (using up $0.03$), then increase the sixth
      and seventh from $0.69$ to $0.72$ (using up $0.06$). With the
      remaining $0.06$ we must still redistribute, we increase the
      sixth, seventh and eighth districts from $0.72$ to $0.74$,
      resulting in
      \begin{equation*}
        (0.37,0.40,0.43,0.46,0.45,0.71,0.71,0.71,0.72,0.75).
      \end{equation*}
    \item \emph{Concentration:} The only district (other than the
      flipped district) whose Democratic vote share is changed is the
      last, resulting in
      \begin{equation*}
        (0.37,0.40,0.43,0.46,0.45,0.63,0.66,0.69,0.72,0.90).
      \end{equation*}
  \end{itemize}
\end{example}

\subsection{Geographic \spc}
\label{sec:sup-geo}

As discussed in the article, we use historical elections in order to
ensure our starting vote distributions are realistic. Our applications
of \spc\ in the article ignore the associated geographic
information. An alternative approach would be to define a
``geographic'' version of \spc\ that physically modified the district
boundaries so as to achieve the desired vote distribution subsequent
to an application of \spc. We have not done this in the article
because we contend that a gerrymandering measure should be able to
record packing and cracking regardless of how feasible the vote
modifications are to the underlying geographic plan.

\begin{figure}
  \centering
  \includegraphics[width=0.7\linewidth]{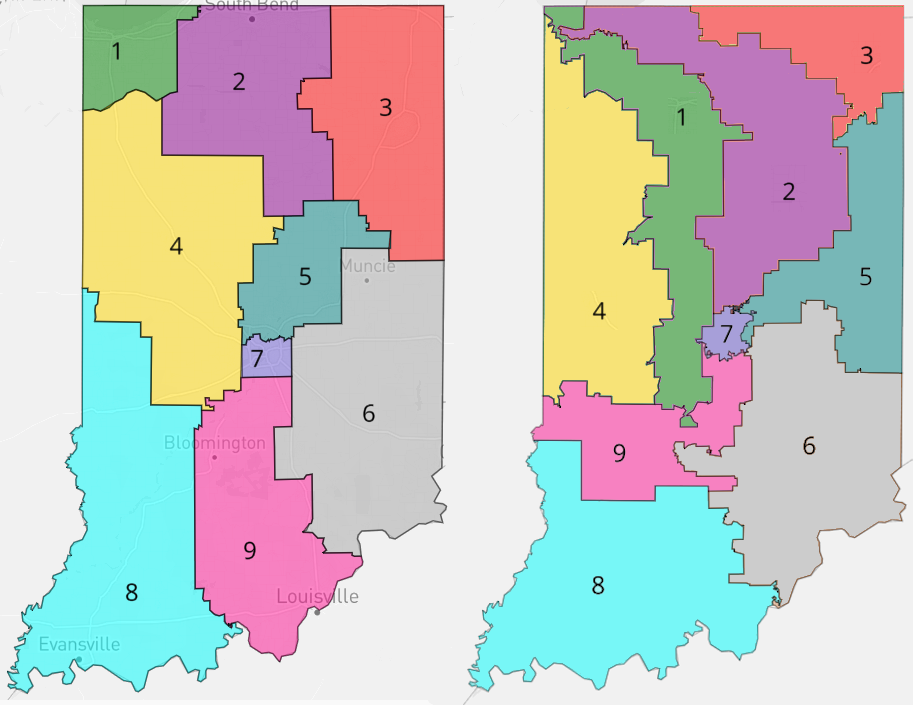}
  \caption{The left image depicts the district plan in place for the
    2018 Indiana House elections. The right image depicts a district
    plan encoding the results of a pro-Republican cracking of one
    district. Both images are from Dave's Redistricting App.}
  \label{fig:geo}
\end{figure}

Nonetheless, it may be helpful to consider a specific example and what
the geographic \spc\ might look like. We consider the Indiana
congressional district plan in place during 2018 elections with the
Democratic vote share in each district determined by a ``partisan
lean'' computed from several statewide races. Taking our data from
Dave's Redistricting App, this yields a vote distribution of
\[ \ell = (0.594, 0.404, 0.340, 0.348, 0.426, 0.334, 0.630, 0.364, 0.387).\]
Districts are numbered as in the left image in Figure~\ref{fig:geo}. A
pro-Republican cracking application of \spc\ requires that we flip
District 1 by redistributing some of its Democratic voters to adjacent
districts in order to give it a Democratic vote share of $0.45$. If we
adhere to the ``Even'' redistribution scheme then this requires
increasing the Democratic vote share in every other district except
District 7 by about $0.02$. This leads to a desired vote distribution
of approximately
\[ \ell^* = (0.450, 0.424, 0.360, 0.368, 0.446, 0.354, 0.630, 0.384, 0.407).\]
Such a district plan\footnote{This district plan was created by hand
in Dave's Redistricting App. The Democratic vote share within each
district is within $0.01$ of the target values in $\ell^*$; the
population of each district is within 0.1\% of the target value. Only
cursory attempts were made to straighten boundaries.} is illustrated
in right image of Figure~\ref{fig:geo}.

The major Democratic areas of the state are Gary in the northwest
(District 1) and Indianapolis in the center (District 7). In order to
flip District 1, we redistributed Democratic voters from the area of
Gary into the neighboring Districts 2 and 4. In order to maintain the
appropriate population and arrive at the desired Democratic vote share
in Districts 2 and 4, we needed to heavily modify their boundaries
with Districts 3, 5, 7, 8 and 9. Similar constraints led to further
modifications of the boundaries among these five districts, as well as
with District 6.

As envisioned, this process places no constraints on the modifications
we make to the various boundaries. The success of geographic
\spc\ reduces to whether or not there exists \emph{any} district plan
with the modified vote distribution of $\ell$. In this case, there
is. As shown in~\citep{Duchin-MA}, sometimes there will not be.

\subsection{Adding noise}
\label{sec:sup-noise}

\begin{table}
  \centering
  \caption{Success of each measure at moving the expected direction
    under one application of \spc\ followed by added Gaussian noise
    with $\sigma=0.025$ for House and state-legislative elections.}
\begin{tabular}{@{}llcccccc@{}}\toprule
  & {} & & \multicolumn{2}{c}{pro-Democratic} & \phantom{ab} & \multicolumn{2}{c}{pro-Republican} \\ \cmidrule{4-5} \cmidrule{7-8}
  & & Fraction correct & Decrease & Increase & & Decrease & Increase \\\midrule
  & \bias & 0.509 & 465 & 471 & & 459 & 500 \\
  & \dec & 0.871 & 828 & 110 & & 135 & 823 \\\rot{\rlap{House}}
  & \eg   & 0.885 & 822 & 116 & & 103 & 857 \\ 
  & \mm   & 0.577 & 531 & 407 & & 395 & 564 \\\midrule
  & \bias & 0.460 & 618 & 672 & & 720 & 569 \\
  & \dec  & 0.621 & 793 & 497 & & 480 & 809 \\
  & \eg   & 0.629 & 822 & 468 & & 489 & 800 \\ \rot{\rlap{Legislative}}
  & \mm   & 0.519 & 691 & 599 & & 642 & 647 \\\bottomrule
\end{tabular}
\label{tab:noise}
\end{table}

Our goal in this article is to explore how well the various measures
we consider record \spc. In doing so, we treat the Democratic vote
share in each district as completely predictable. We have therefore
set up a framework that should be \emph{easier} than the real world.

\begin{figure}
  \centering
  \includegraphics[width=1\linewidth]{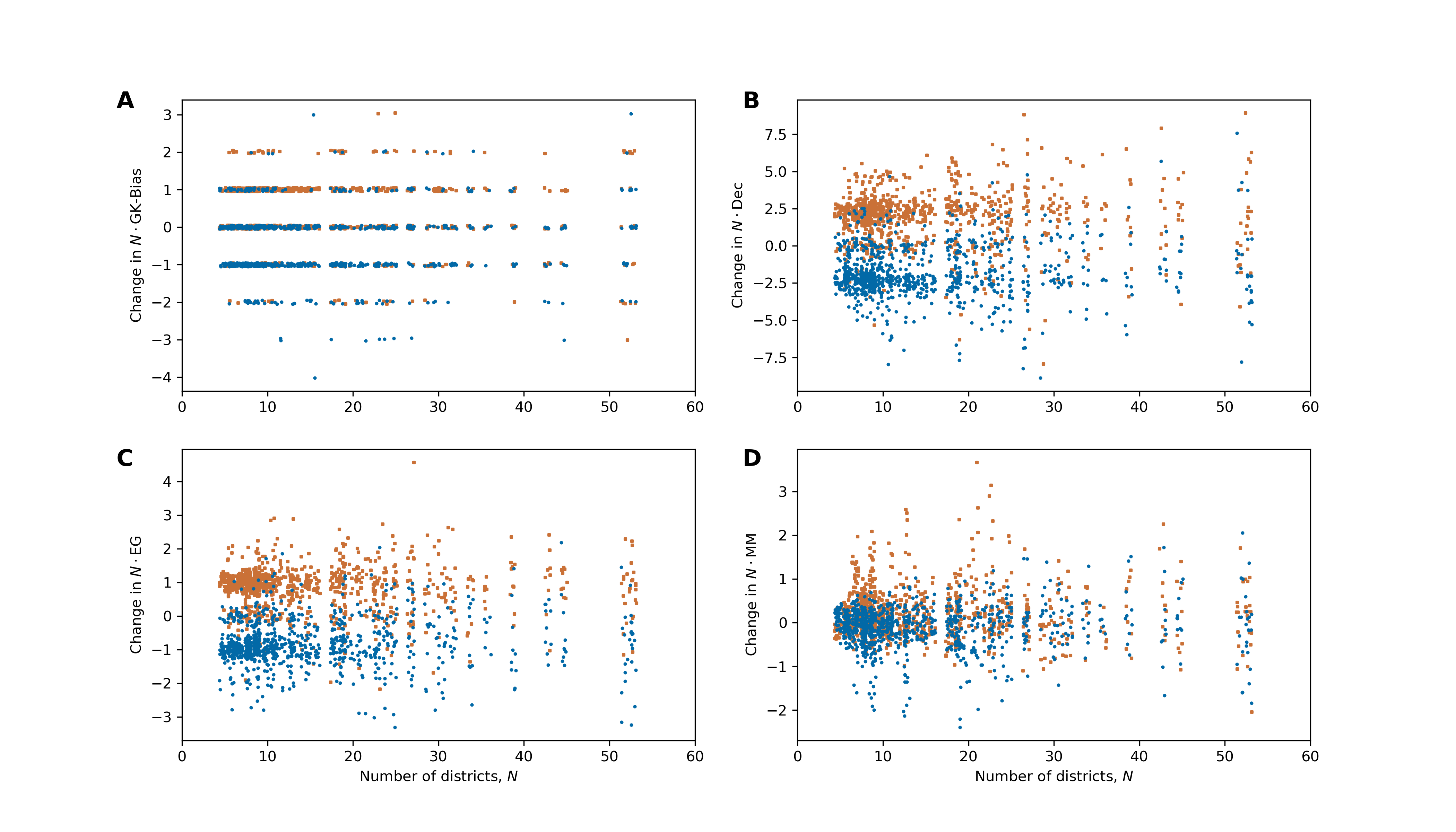}
  \caption{Plot of change in measures (each scaled by $N$) due to \spc\ 
    followed by added Gaussian noise with $\sigma=0.025$. For each
    subplot, the red squares represent single applications of
    pro-Republican \spc\ to House elections since 1972; the blue
    circles represent single applications of pro-Democratic \spc.
    Measures depicted are partisan bias (A), declination (B),
    efficiency gap (C) and mean-median difference (D). Horizontal and
    vertical coordinates have been jittered to reduce overlap.  }
  \label{fig:noise}
\end{figure}

A disadvantage of this approach is that it might lead to false
confidence regarding those measures that do well under our model.  To
take a concrete example, consider the election $\ell =
(0.2,0.501,0.8)$. If we apply pro-Republican cracking as in our
default version of \spc, we arrive at a resulting distribution of
$\ell^* = (0.251,0.45,0.8)$. The efficiency gap measure perfectly
records the flipping of one seat with $N\cdot \eg$ changing from
$-1/2$ to $+1/2$, leading to a change of $+1$. However, the middle
district in election $\ell$ is without question a toss
up. Intuitively, the Democrats will win two of the three seats only
half the time for the original election $\ell$ (and consistently win
only one seat in the modified distribution $\ell^*$). So, the
\emph{average} change in $N\cdot \eg$ in the real, noisy world should
be only $+1/2$, not $+1$.

With the above example in mind, we include here data that will
hopefully provide some insight into how each measure might perform in
an environment in which there is more unpredictably. For each
election, we apply \spc\ as described in the main body of the
paper. We then add Gaussian noise from a normal distribution with
standard deviation $\sigma = 0.025$ to the Democratic vote share in
each district for the starting distribution as well as for the
distribution resulting from the application of \spc.\footnote{With
this value, 95\% of the time the noise will not affect the winner in
``uncompetitive'' districts (those in which the Democratic vote share
is either less than 0.45 or more than 0.55).}  Continuing our example
from above, one trial changes $\ell$ to $\tilde{\ell} =
[0.228,0.498,0.797]$ and $\ell^*$ to $\tilde{\ell}^* =
[0.251,0.478,0.751]$. In this instance, the middle district in $\ell$
does switch to Republican with the added noise, resulting in no change
in the value of efficiency gap between $\tilde{\ell}$ and
$\tilde{\ell}^*$.

We revisit Table~\ref{tab:fits} and
Figure~\ref{fig:pandc-1-mpandc-11}, but now with added noise as
described above. The results for $\sigma=0.025$ are presented in
Table~\ref{tab:noise} and~Figure~\ref{fig:noise}. Results for each
application of \spc\ are averaged over ten trials (i.e., we follow the
procedure described in the previous paragraph ten times and average
the change). Comparing to the data in Table~\ref{tab:fits}, we see
that for House elections, the declination and efficiency gap continue
to do quite well, partisan bias does about the same, and mean-median
difference even does a little better. For the legislative elections,
none do particularly well. This is presumably connected to the fact
that, with the large number of seats typically found in a legislative
election, there are frequently multiple competitive districts that are
susceptible to being affected by the added noise.

\end{document}